\documentclass[12pt]{iopart}

\usepackage{graphicx}

\begin{document}

\title[Magnetic topological Weyl fermions in half-metallic In$_2$CoSe$_4$]{Magnetic topological Weyl fermions in half-metallic In$_2$CoSe$_4$}

\author{Xiaosong Bai, Yan Wang, Wenwen Yang, Qiunan Xu$^{*}$, Wenjian Liu}
\address{Qingdao Institute for Theoretical and Computational Sciences, School of Chemistry and Chemical Engineering, Shandong University, Qingdao 266237, China}
\ead{xuqiunan91@email.sdu.edu.cn}

\vspace{10pt}
\begin{indented}
\item[]September 2024
\end{indented}

\begin{abstract}
Magnetic Weyl semimetals (WSM) have recently attracted much attention
due to their potential in realizing strong anomalous Hall effects.
Yet, how to design such systems remains unclear.
Based on first-principles calculations,
we show here that the ferromagnetic half-metallic compound In$_2$CoSe$_4$ has several pairs of Weyl points
and is hence a good candidate for magnetic WSM.
These Weyl points would approach the Fermi level gradually as the Hubbard $U$ increases,
and finally disappear after a critical value $U_c$.
The range of the Hubbard $U$ that can realize the magnetic WSM state can be expanded by pressure,
manifesting the practical utility of the present prediction.
Moreover, by generating two surface terminations at Co or In atom
after cleaving the compound at the Co-Se bonds,
the nontrivial Fermi arcs connecting one pair of Weyl points with opposite chirality are discovered in surface states.
Furthermore, it is possible to observe the nontrivial surface state experimentally,
e.g., angle-resolved photoemission spectroscopy (ARPES) measurements.
As such, the present findings imply strongly a new magnetic WSM
which may host a large anomalous Hall conductivity.
\end{abstract}

%
\vspace{2pc}

\submitto{\JPCM}
%
\maketitle
%
%

\section{Introduction}

The principles of topology receive extensive attention along with the discovery of materials
with exotic physical and chemical properties \cite{Hasan2010ku,Qi2011RMP,zhang2009,Wan2011,Burkov2011,Young2012,Wang2012,Weng2015,Yan2017,science.aaf5037,TM1,TM2,TM3}.
Weyl semimetal (WSM) is one of such classes of topological materials whose
Berry curvatures exhibit magnetic monopoles in momentum space \cite{science1089408}.
The doubly degenerate Weyl fermions, which are formed by the linear cross of conduction and valence bands,
always appear in pairs with opposite chirality, and can be annihilated if this pair of Weyl points overlap with each other.
Similar to topological insulators, WSMs can also exhibit nontrivial surface states,
which connect one pair of Weyl points with different chirality,
and appear as a non-closed Fermi arc in momentum space.
The Fermi arcs provide the most intuitive and significant feature for the measurement and verification of WSMs.
Moreover, the existence of Weyl fermions and Fermi arcs brings about many novel transport properties,
such as chiral anomaly effect \cite{HOSUR2013857,Huang2015anomaly,Zhang2016ABJ,RevModPhys.90.015001},
strong anomalous Hall effect \cite{Burkov:2011de,Xu2011,CoSnS-NP}, and
giant magnetoresistance \cite{Shekhar2015,Ghimire2015}.

In solids, Weyl fermions can usually be formed by breaking space inversion symmetry or time reversal symmetry.
The WSMs that break space inversion symmetry have been theoretically predicted and experimentally verified
in type-I WSM non-centrosymmetric transition-metal monophosphides (Ta/Nb)(As/P) \cite{Weng2015,Hasan2015,Huang2015anomaly,PhysRevX.5.031013,science.aaa9297}, and
type-II WSM (W/Mo)Te$_2$ \cite{nature15768,PhysRevB.92.161107,nphys3871}, etc.
Likewise, the WSMs that break time reversal symmetry have also been theoretically predicted and experimentally verified
in ferromagnetic (FM) Kagom\'{e} crystal Co$_3$Sn$_2$S$_2$ \cite{CoSnS-NP,PhysRevB.97.235416,CoSnS-Weng,science.aav2873,science.aav2334}
and FM ordered state of EuB$_6$ \cite{PhysRevLett.124.076403,PhysRevX.11.021016}.
However, the number of predicted magnetic WSM candidates is far less than nonmagnetic WSMs
because of the application of high-throughput calculations in nonmagnetic system \cite{TM1,TM2,TM3,HT-xu,GAO2021667}
and the difficulty about observing topological phase in magnetic materials.
Compared to nonmagnetic WSMs, magnetic WSMs can host the fewest Weyl fermions,
which is beneficial for the exploration of novel physical mechanisms.
Meanwhile, the quantum anomalous Hall effect with intrinsic magnetism can be realized in magnetic systems exactly,
and the combination of topology and spin of electrons may push forward the development of spintronics.

In this paper, based on first-principles calculations, we demonstrate that the FM
In$_2$CoSe$_4$ is a potential candidate for magnetic WSM.
First of all, by means of total energy calculations,
the ground state for magnetic In$_2$CoSe$_4$ is found to be FM
with magnetic moments of Co atoms ordered along the [001] or [111] directions.
Secondly, eight or twelve independent Weyl points near the Fermi level are observed
for the two directions of magnetic moments, respectively,
when the Hubbard $U$ of Co atom is smaller than the threshold $U_c$.
The Weyl points will be annihilated if we further increase the Hubbard $U$.
The critical value $U_c$ of the topological transition can be adjusted by the effect of pressure,
making the realization of Weyl fermions possible in experiment.
Finally, the surface states of FM In$_2$CoSe$_4$ show that
the (001) surface states with magnetic moment point along [111] direction are clearly and observably,
which can readily be measured experimentally.

\section{Computational Methods}
The electronic and magnetic structures of In$_2$CoSe$_4$ were investigated by using density functional theory (DFT),
with the Perdew-Burke-Ernzerhof (PBE) variant\cite{perdew1996} of the
generalized gradient approximation to the exchange-correlation density functional,
as implemented in the Vienna Ab-initio Simulation Package (VASP)\cite{kresse1996}.
The energy cutoff for the plane wave basis was chosen to be 400 eV and
the k-mesh in the self-consistent process was set to be 8$\times$8$\times$8.
The experimental lattice constants of In$_2$CoSe$_4$ were taken from Ref. \cite{icsd658918},
and were fully relaxed until the force on each atom was less than 1$\times$10$^{-3}$ eV/\AA.
The DFT+U approach\cite{PhysRevB.57.1505} was employed to describe the 3d orbitals of Co atoms
with the Hubbard $U$ changing from 0 to 2 eV.
In order to investigate the surface states of the (001) and (100) surfaces,
a tight-binding model, which was composed of In 5s5p, Co 4d and Se 4p,
was constructed in terms of maximally localized Wannier functions (MLWFs)\cite{Mostofi2008}.
The Green's function method\cite{Sancho1984,Sancho1985} was then employed to calculate
the surface states under the half-infinite boundary conditions.
Furthermore, the nodal lines and Weyl points with non-zero chirality were located by the energy difference of band structures,
which was derived from the tight-binding Hamiltonian.

\section{Results and discussion}
\begin{figure*}[htb]
\centering
\includegraphics[width=1.0\textwidth]{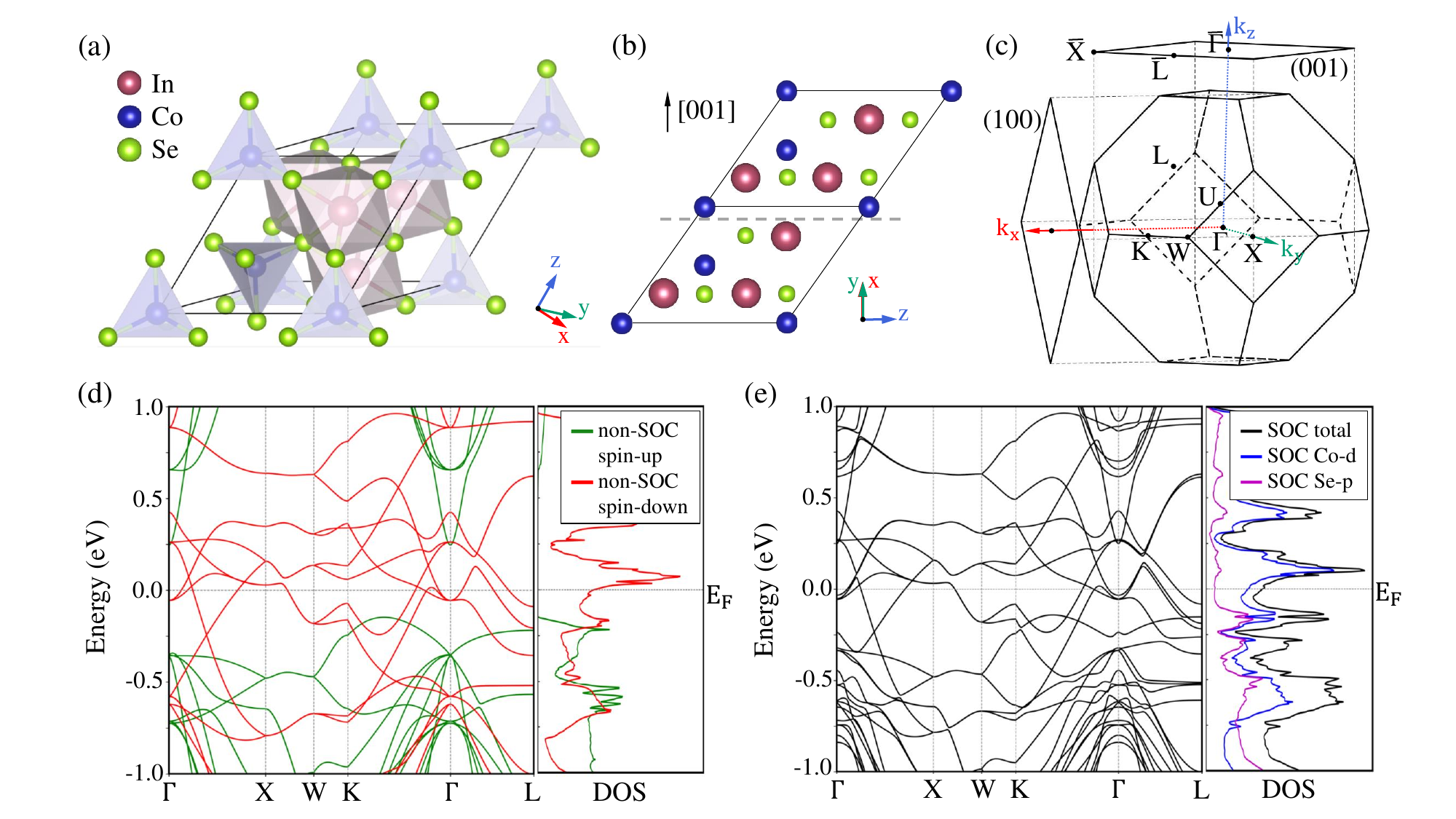}
   \caption{
Crystal structure and electronic structure of ferromagnetic (FM) In$_2$CoSe$_4$.
(a) Unit cell of In$_2$CoSe$_4$.
(b) Side view of the crystal structure. The dashed line represents the cross section to establish the (001) surfaces with Co-termination and In-termination.
(c) First Brillouin zone of space group Fm$\bar{3}$m and surface Brillouin zones of (001) and (100) surfaces. High symmetry k-points are labeled.
(d) Band structure and density of states (DOS) for FM In$_2$CoSe$_4$ without spin-orbital coupling (SOC) when Hubbard $U$ = 0 eV, which show half-metallic features. Red lines and green lines are spin up and spin down, respectively.
(e) Band structure, total DOS and partial DOS for FM In$_2$CoSe$_4$ with SOC when Hubbard $U$ = 0 eV and magnetic moment $\mu \mathop{//} [001]$, which show that the inverted bands are composed by 3d orbits of Co atoms.
}
\label{fig1}
\end{figure*}

\subsection{Crystal structure and magnetism}

The crystal structure of In$_2$CoSe$_4$ is cubic with nonsymmorphic space group
Fd$\overline{3}$m (No.227) \cite{icsd658918}, just like silicon.
It has inversion center, two-fold rotation axes $2_{001/010/100}$, three-fold rotation
axis $\pm 3^{\pm}_{111}$, mirror planes $m_{1-10/01-1/-101}$, slip planes, and spiral axes.
The optimized structure is in very good agreement with experiment, as shown in Table~\ref{tab1},
in which the magnetic Co atom and four equivalent Se atoms form a CoSe$_4$ tetrahedron,
and In atom and six equivalent Se atoms form a InSe$_6$ octahedron.
All InSe$_6$ octahedra share corners with six CoSe$_4$ tetrahedra and edges with six InSe$_6$ octahedra,
which is shown in Fig.~\ref{fig1}(a).
This structure can also be viewed as a stack of Co layers and In-Se layers alternately along
[001] direction of fractional coordinate system, as shown in Fig.~\ref{fig1}(b).

\begin{table}
\caption{\label{tab1}Experimental and theoretical lattice constants for the conventional cell of In$_2$CoSe$_4$}
\footnotesize\rm
\begin{tabular*}{\textwidth}{@{}l*{15}{@{\extracolsep{0pt plus12pt}}l}}
\br
    & Ref. \cite{icsd658918}   & optimized \\
\mr
$a$ [\AA]                           & 11.05                 &  11.14\\
$\alpha$ [$^{\circ}$]               & 90                    &  90\\
Co                                  & [1/4, 3/4, 1/4] &  [1/4, 3/4, 1/4]\\
In                                  & [5/8, 7/8, 3/8] &  [5/8, 7/8, 3/8]\\
Se                                  & [0.871, 0.129, 0.129] &  [0.8749, 0.1251, 0.1251] \\
\br
\end{tabular*}
\end{table}

As of now, the magnetic structure of In$_2$CoSe$_4$ has not been studied in detail by both experiment and theory.
Therefore, we examined all possible collinear magnetic orders
for In$_2$CoSe$_4$, including paramagnetism, ferromagnetism, and antiferromagnetism.
Based on the calculated total energy, the ground state of In$_2$CoSe$_4$ is determined to be FM.
We also looked at different orientations of the magnetic moments of Co atoms in FM In$_2$CoSe$_4$,
and found that the total energy is lowest when the magnetic moments point to the [001] or [111] direction,
with the difference in between being less than 10$^{-4}$ eV.
Thus, in the following discussion of Weyl fermions,
we will mainly focus on the FM In$_2$CoSe$_4$ with the magnetic moments
of Co atoms pointing along the [001] and [111] directions.

\subsection{Influence of the Hubbard $U$}

The electronic structure of FM In$_2$CoSe$_4$ with zero Hubbard $U$ is shown in Fig.~\ref{fig1}(d) and~\ref{fig1}(e),
where the Fermi energy is set to 0.
In the absence of spin-orbit coupling (SOC), FM In$_2$CoSe$_4$ is half-metallic, that is,
the spin-up channel (green lines) is insulated with a band gap as large as 387.1 meV,
whereas the spin-down channel (red lines) is metallic with nodal line structure near the Fermi level.
When the SOC is taken into account, the nodal line structure would open a gap
to form Weyl points due to the breaking of time reversal symmetry.
Since the band structures in the presence of SOC are similar for
the magnetic moments $\mu$ pointing to the [001] and [111] directions,
we just show the electronic structure with magnetic moment $\mu \mathop{//} [001]$ in Fig.~\ref{fig1}(e).
According to the partial density of states,
the band inversion around $\Gamma$ point is composed of the 3d orbits of Co atoms.

\begin{figure*}[htb]
\centering
\includegraphics[width=1.0\textwidth]{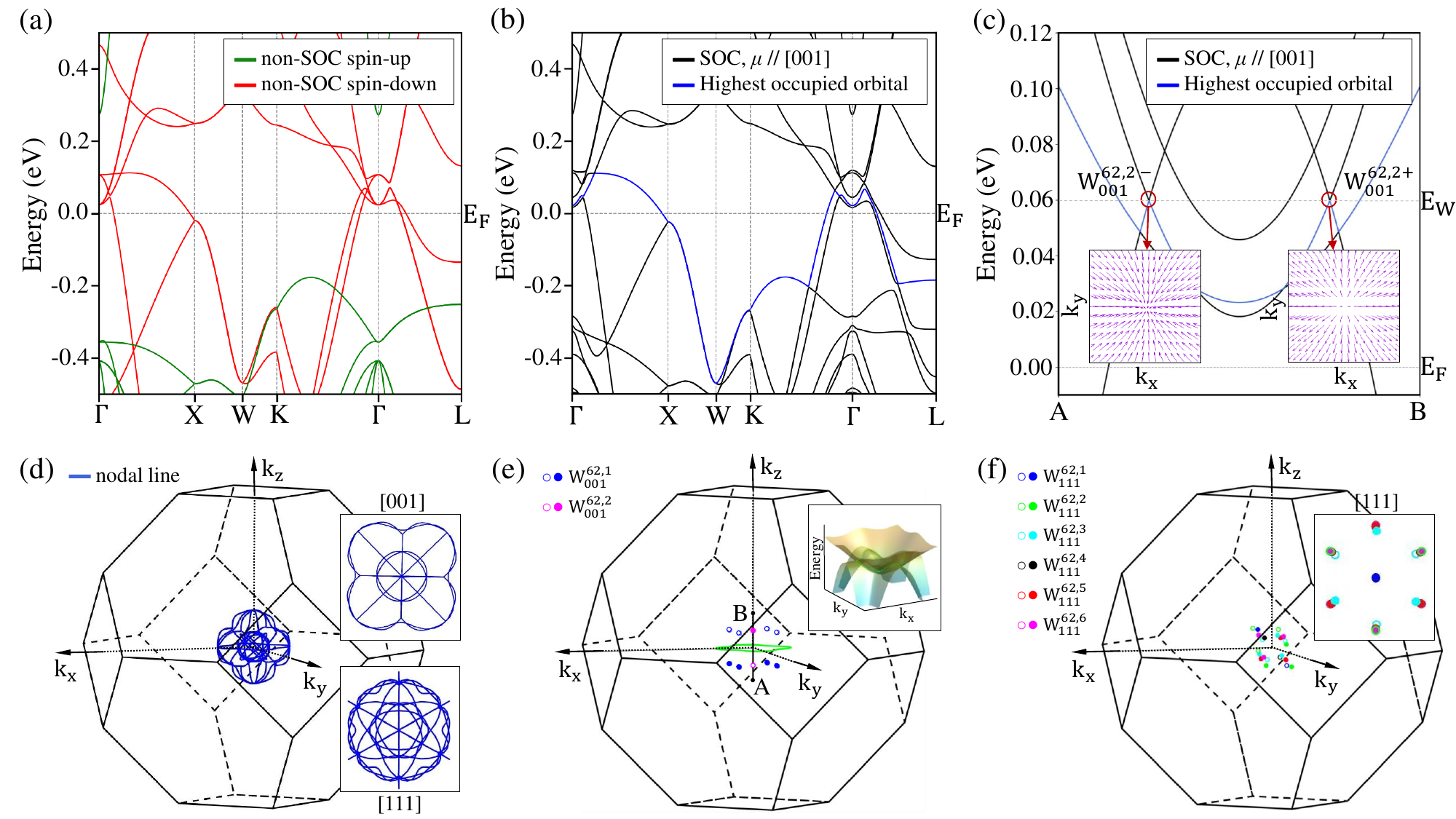}
   \caption{
Weyl points of FM In$_2$CoSe$_4$ with Hubbard $U$ = 1 eV.
(a) Band structure without SOC.
(b) Band structure with SOC when magnetic moment $\mu$ point to [001] direction.
(c) Energy dispersion along the k-path $AB$ crossing a pair of Weyl points $W_{001}^{62,2+}$ and $W_{001}^{62,2-}$ as shown in (e). $E_F$ and $E_W$ are the energy of Fermi level and Weyl points. The inserted illustrations are the Berry curvature for Weyl points.
The blue line in (b) and (c) represents the highest occupied band with band number 62.
(d) Nodal line structure when SOC is not considered. Two insets show the top view of nodal line structure along [001] and [111] directions, respectively.
(e) and (f) Location of Weyl points when SOC is considered for $\mu \mathop{//} [001]$ and $\mu \mathop{//} [111]$, respectively. The small picture in (e) are the three-dimensional dispersion of nodal line structure with small gap, and the small picture in (f) is the top view along the [111] direction. The solid and hollow circle represent different chirality, respectively.
}
\label{fig2}
\end{figure*}

To see how the topological properties of FM In$_2$CoSe$_4$ vary with respect to the 3d orbits of Co atoms,
we carried out calculations with different Hubbard $U$ for the Co 3d orbits.
When $U$ is increased from 0 to 2 eV,
the magnetic moment for each Co atom increases from 2.14 to 2.35 $\mu_B$,
and the electronic structure changes from metallic state to insulating state.
At the same time, the Weyl points with different chirality in metallic state gather together to $\Gamma$ point
and annihilate with each other thoroughly to form a trivial state.
That is to say, the FM In$_2$CoSe$_4$ will be transformed from a magnetic Weyl semimetal
into a trivial insulator when the Hubbard $U$ is larger than the threshold $U_c$ = 1.53 eV,
which is much smaller than the empiric value for Co atoms.
In general, the actual value of $U$ depends on the chemical environment of the magnetic Co atoms.
In order to make it possible to realize the magnetic WSM state in FM In$_2$CoSe$_4$ in experiment,
we can judge the threshold $U_c$ for topological transition by imposing external forces.
For example, we find that $U_c$ will increase when the pressure is applied.
The DFT calculation indicate that the critical point $U_c$ will be 2.68 eV when the lattice constant is reduced by 4.0 $\%$.
Moreover, the band structure with $U$ = 1 eV has less trivial states near Fermi level as shown in Fig.~\ref{fig2}(a) and~\ref{fig2}(b),
which can exhibit the topological properties more clearly.
Thus, we take $U$ = 1 eV as example in the subsequent discussions on Weyl fermions and topological surface states.

\subsection{Weyl fermions of FM WSM In$_2$CoSe$_4$}

According to the band structure of FM In$_2$CoSe$_4$ with Hubbard $U$ = 1 eV in the absence of SOC in Fig.~\ref{fig2}(a),
five bands for spin-down near the Fermi level will form nodal lines as shown in Fig.~\ref{fig2}(d).
All the nodal lines connect with each other to compose a cage around the $\Gamma$ point of first Brillouin zone.
After taking the SOC effect into account, the nodal line structure will open most of the gaps but leave several
isolated doubly degenerate points, i.e., Weyl points with chirality $\pm1$.
For convenient, the two bands form nodal lines and Weyl points are labeled as n and n+1, where the band number n goes from 61 to 64.
Thus, the independent Weyl points within the energy range of -0.1 eV $\sim$ 0.1 eV are listed in Table~\ref{tab2},
where the subscript of the mark W represents the direction of magnetic moment,
and the superscript represents the combination of n and serial number.
The band structures with SOC in case of $\mu \mathop{//} [001]$ are shown in Fig.~\ref{fig2}(b) and~\ref{fig2}(c),
which walk along the k-path same with non-SOC case and crossing one pair of Weyl points $W_{001}^{62,2\pm}$ at the $k_z$ axis, respectively.
The inserted illustrations in Fig.~\ref{fig2}(c) are the sink and the source of Berry curvature for Weyl points with opposite chirality.

\begin{table}
\caption{\label{tab2}Location of independent Weyl points for FM In$_2$CoSe$_4$ with magnetic moments pointing to [001] and [111] directions in the energy range of -0.1 eV $\sim$ 0.1 eV.}
\footnotesize\rm
\begin{tabular*}{\textwidth}{@{}l*{15}{@{\extracolsep{0pt plus12pt}}l}}
\br
WP & cartesian coordinate & chirality & number & energy/eV  \\
\mr
$W_{001}^{61,1}$ & $[\ \ 0.0000,\ \ 0.0000,\ \ 0.0332]$ & $+1$  &  2  &  0.0276\\
$W_{001}^{61,2}$ & $[\ \ 0.0553,\ \ 0.0553,\ \ 0.0676]$ & $-1$  &  8  &  0.0317\\
$W_{001}^{62,1}$ & $[\ \ 0.0704,\ \ 0.0704,\ \ 0.0611]$ & $+1$  &  8  &  0.0355\\
$W_{001}^{62,2}$ & $[\ \ 0.0000,\ \ 0.0000,\ \ 0.0630]$ & $-1$  &  2  &  0.0572\\
$W_{001}^{63,1}$ & $[\ \ 0.0000,\ \ 0.0000,\ \ 0.0516]$ & $+1$  &  2  &  0.0752\\
$W_{001}^{63,2}$ & $[\ \ 0.0000,\ \ 0.0410,\ \ 0.0441]$ & $-1$  &  8  &  0.0785\\
$W_{001}^{63,3}$ & $[\ \ 0.0411,\ \ 0.0411,\ \ 0.0362]$ & $+1$  &  8  &  0.0851\\
$W_{001}^{63,4}$ & $[\ \ 0.0000,\ \ 0.0747,\ \ 0.0294]$ & $+1$  &  8  &  0.0955\\
\mr
$W_{111}^{61,1}$ & $[\ \ 0.0257,\ \ 0.0257,\ \ 0.0257]$ & $+1$  &  2  &  0.0359\\
$W_{111}^{61,2}$ & $[\ \ 0.0559,\ \ 0.0559,\ \ 0.0559]$ & $+1$  &  2  &  0.0394\\
$W_{111}^{61,3}$ & $[\ \ 0.0368,\ \ 0.0576,\ \ 0.0576]$ & $-1$  &  6  &  0.0430\\
$W_{111}^{62,1}$ & $[\ \ 0.0711,\ \ 0.0711,\ \ 0.0711]$ & $-1$  &  2  &  0.0303\\
$W_{111}^{62,2}$ & $[-0.0406,\ \ 0.0245,\ \ 0.0245]$ & $+1$  &  6  &  0.0642\\
$W_{111}^{62,3}$ & $[-0.0045,\ \ 0.0694,\ \ 0.0694]$ & $+1$  &  6  &  0.0646\\
$W_{111}^{62,4}$ & $[\ \ 0.0387,\ \ 0.0387,\ \ 0.0387]$ & $-1$  &  2  &  0.0654\\
$W_{111}^{62,5}$ & $[-0.0305,\ \ 0.0405,\ \ 0.0405]$ & $+1$  &  6  &  0.0744\\
$W_{111}^{62,6}$ & $[-0.0311,\ \ 0.0443,\ \ 0.0443]$ & $+1$  &  6  &  0.0756\\
$W_{111}^{63,1}$ & $[-0.0153,\ \ 0.0471,\ \ 0.0471]$ & $-1$  &  6  &  0.0790\\
$W_{111}^{63,2}$ & $[\ \ 0.0151,\ \ 0.0531,\ \ 0.0531]$ & $+1$  &  6  &  0.0823\\
$W_{111}^{63,3}$ & $[-0.0703,\ \ 0.0241,\ \ 0.0241]$ & $+1$  &  6  &  0.0999\\
\br
\end{tabular*}
\end{table}

For a better illustration of the Weyl points in the hole Brillouin zone,
we show the Weyl points with two different orientations of magnetic moments in Fig.~\ref{fig2}(e) and~\ref{fig2}(f)
when n is the highest occupied band 62.
In the case of the FM magnetic moment $\mu \mathop{//} [001]$,
the magnetic structure of In$_2$CoSe$_4$ maintains inversion symmetry,
two-fold rotation symmetry $2_{001}$, four-fold rotation symmetry $4^{\pm}_{001}$ and mirror symmetry $m_{001}$,
but not the three-fold rotation symmetry.
Two independent Weyl points denoted as $W_{001}^{62,1-2}$ remain in the end.
The one located at the $k_z$ axis has one pair of Weyl points
and the other one have four pairs as shown in Fig.~\ref{fig2}(c).
At the same time, a small-gapped nodal line in the $k_z$ = 0 plane is also left over,
which is labeled by green circle, with a negligible gap below 0.3 meV.
Moreover, in the case of the FM magnetic moment $\mu \mathop{//} [111]$,
the magnetic structure of In$_2$CoSe$_4$ only has inversion symmetry and three-fold rotation symmetry $\pm 3^{\pm}_{111}$.
There exist six independent Weyl points denoted as $W_{111}^{62,1-6}$, among which
the Weyl points $W_{111}^{62,1}$ and $W_{111}^{62,2}$ on the [111] axis have only one pair,
and the other four Weyl points around the [111] axis have three pairs.

\subsection{Surface states of FM WSM In$_2$CoSe$_4$}

For WSMs, the surface state is shown as a non-closed Fermi arc which connects one pair of Weyl points with opposite chirality.
In order to investigate the nontrivial surface states, we projected the band structure of semi-infinite In$_2$CoSe$_4$
with magnetic moments $\mu \mathop{//} [001]$ and $\mu \mathop{//} [111]$ onto the (100) and (001) surfaces, respectively.
The cleavage surface will break Co-Se bonds which is shown by the dashed line in Fig.~\ref{fig1}(b),
and form two different terminations with Co or In-Se layers.

\begin{figure*}[htb]
\centering
\includegraphics[width=0.8\textwidth]{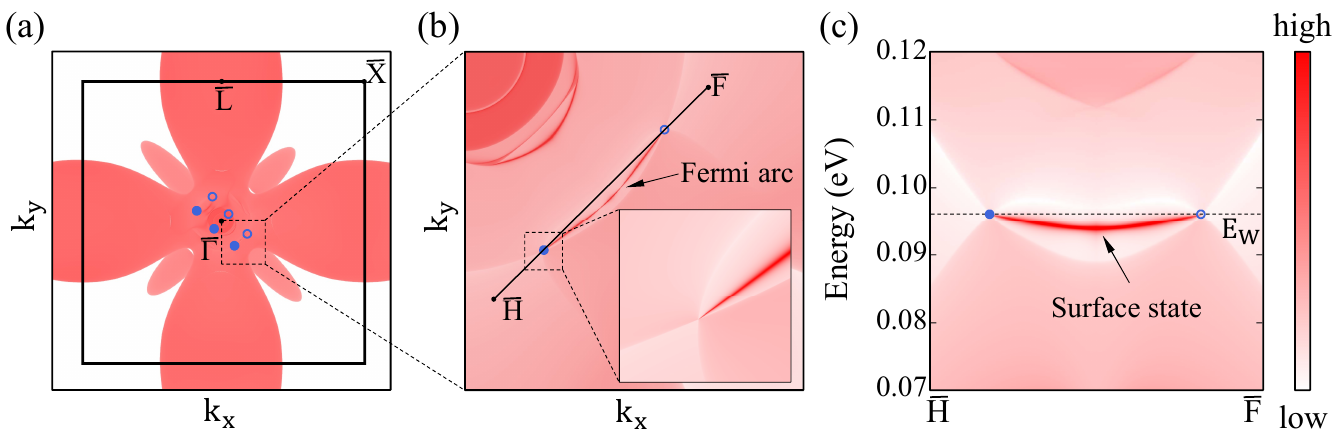}
   \caption{
(100) surface states of FM In$_2$CoSe$_4$ with magnetic moment $\mu \mathop{//} [001]$ for In-terminated surface.
(a) Fermi arcs with energy fixed at the Weyl points $W_{001}^{62,1}$.
(b) Zoom-in view of the dotted box in (a).
(c) Energy dispersion along the k-path $\bar{H}\bar{F}$ crossing one pair of Weyl points which are connected by a Fermi arc.
}
\label{fig3}
\end{figure*}

\begin{figure*}[htb]
\centering
\includegraphics[width=1.0\textwidth]{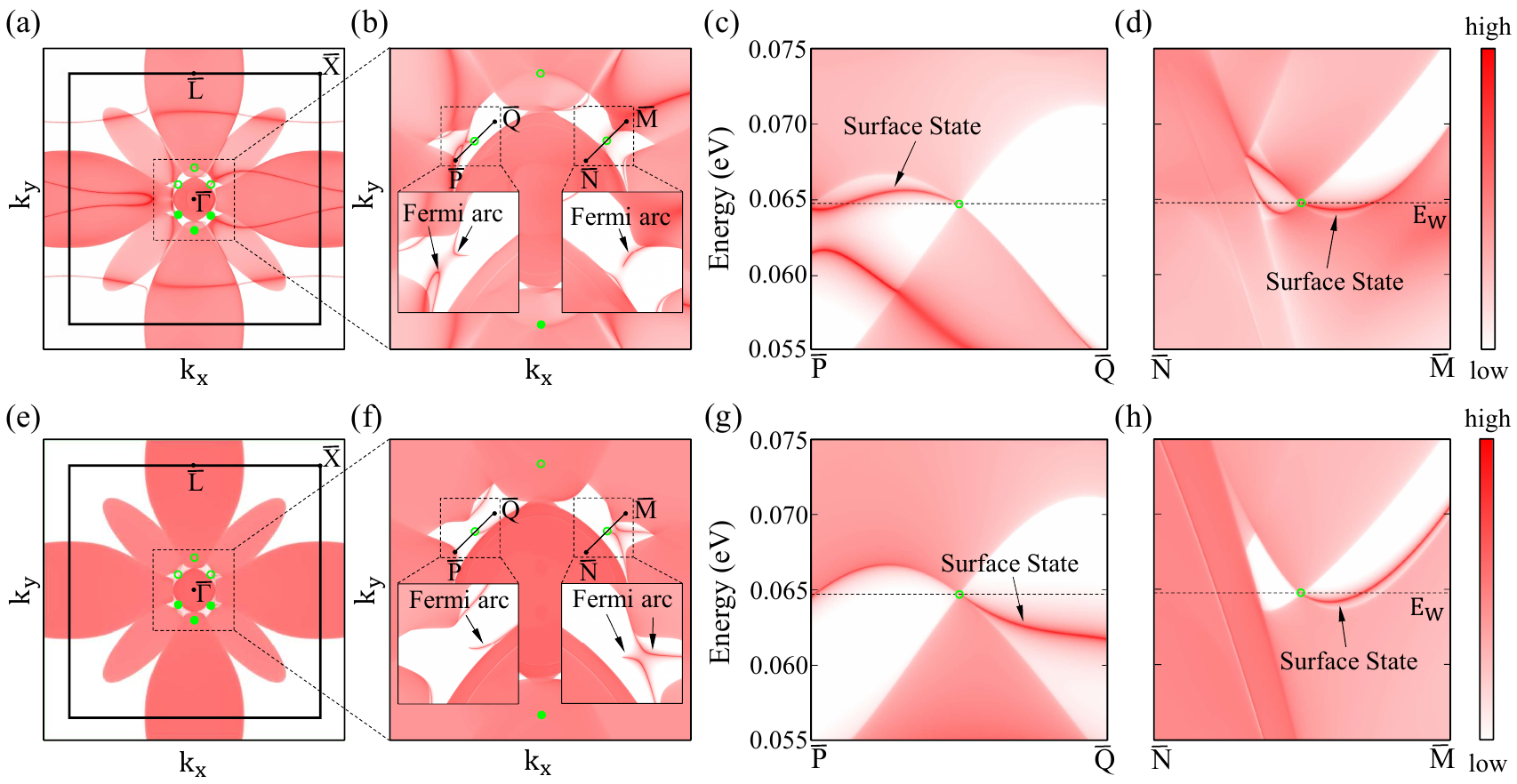}
   \caption{
(001) surface states of FM In$_2$CoSe$_4$ with magnetic moment $\mu \mathop{//} [111]$ for Co-terminated and In-terminated surface.
(a)-(b) Fermi arcs for Co-terminated surface with energy fixed at the Weyl points $W_{111}^{62,2}$. (b) is the zoom-in view of central part of (a).
(c)-(d) Energy dispersion for Co-terminated surface crossing Weyl point along $\bar{P}\bar{Q}$ and $\bar{N}\bar{M}$, respectively.
(e)-(f) Fermi arcs for In-terminated surface with energy fixed at the Weyl points $W_{111}^{62,2}$. (f) is the zoom-in view of central part of (e).
(g)-(h) Energy dispersion for In-terminated surface crossing Weyl point along $\bar{P}\bar{Q}$ and $\bar{N}\bar{M}$, respectively.
The color bar for surface states demonstrates the intensity of states, and red/white represents the occupied/unoccupied states.
}
\label{fig4}
\end{figure*}

When the magnetic moment points along [001] direction,
the mirror symmetry will make two Weyl points with opposite chirality overlap with each other in (001) surface as shown in Fig.~\ref{fig2}(e).
Therefore, no Fermi arcs can be observed in (001) surface.
In order to observe the Fermi arcs evidently, we calculated the projected band structure of (100) surface.
By using Green's function method, we found that the Weyl points and nontrivial surface states are all merged into the bulk state
of FM In$_2$CoSe$_4$ when $\mu \mathop{//} [001]$.
For instance, the (100) surface Fermi surface (FS) for In-termination with energy fixed at the Weyl points $W_{001}^{62,1}$,
which is shown in Fig.~\ref{fig3}(a), would not be observed visually unless the color bar is judged.
The zoom-in view around one pair of projected Weyl points with opposite chirality is shown in Fig.~\ref{fig3}(b),
indicating that the Fermi arc connects these two Weyl points directly.
The nearly straight Fermi arc indicate that FM In$_2$CoSe$_4$ may host high surface conductivity 
since the straight arc is more disorder tolerant than regular ones \cite{surfcond}.
Meanwhile, we investigated the band dispersion along the k-path crossing this pair of type-II Weyl points in Fig.~\ref{fig3}(c),
where the surface state is nearly flat with slight broadening.
However, the (100) surface states of $W_{001}^{62,1}$ are not distinguishable, although the topological properties for WSMs can be achieved perfectly.

When the magnetic moment points along the [111] direction,
the (001) surface states corresponding to two different terminations can readily be observed for the Weyl points $W_{111}^{62,2}$.
According to the surface FS with the energy fixed at $W_{111}^{62,2}$ (see Fig.~\ref{fig4}(a) and (e)),
in total three pairs of projected Weyl points can be observed around $\Gamma$ point,
but one pair of them along $k_y$ axis merges into the bulk state.
The non-closed Fermi arc surface state starting from Weyl point can be observed in the zoom-in view of FSs,
which is shown in Fig.~\ref{fig4}(b) and (f).
In order to check for the possible nontrivial Fermi arcs,
we also analyzed the band dispersion along the k-path crossing the Weyl points $\bar{P}\bar{Q}$ and $\bar{N}\bar{M}$.
According to the surface state dispersion for Co-terminated surface as shown in Fig.~\ref{fig4}(c),
it can be noted that the arc at the bottom left of the Fermi arc in $\bar{P}\bar{Q}$ is also a nontrivial state.
Meanwhile, the arc at the upper right of the Fermi arc in$\bar{N}\bar{M}$ is also nontrivial,
in accordance with the dispersion for the In-terminated surface as indicated in Fig.~\ref{fig4}(h).

\section{Conclusions}

In summary, we have theoretically predicted a ferromagnetic topological WSM In$_2$CoSe$_4$
with the magnetic moment points along the [001] or [111] direction.
The location of these Weyl points will change slightly with the increase of the Hubbard $U$.
The Weyl points will disappear if $U$ is larger than the critical value $U_c$ = 1.53 eV.
The critical Hubbard $U_c$ increases remarkably when the pressure is applied.
When the SOC is neglected, FM In$_2$CoSe$_4$ is a half-metal whose spin-down channel can host a list of nodal lines.
When the SOC is taken into account, the nodal lines will open a gap.
Consequently, 8 or 12 independent Weyl points can be observed near the Fermi level
when the magnetic moments point along the [001] and [111] directions, respectively.
In order to study the surface states of FM WSM In$_2$CoSe$_4$,
we calculated the isoenergetic surface with the energy fixed at the Weyl point and the band dispersion
based on a half-infinite system with one surface.
Most surface states are submerged into the bulk state except the (001) surface states
in the case of the magnetic moment $\mu \mathop{//} [111]$ for the Co-terminated and In-terminated surfaces.
The Fermi arcs connecting the Weyl points $W_{111}^{62,2}$ with opposite chirality can be observed clearly.
It is believed that the present systematic investigations are helpful for future experimental fabrication.

\ack{The work was supported by the National Key R\&D Program of China (Grants No. 2023YFA1506901),
the National Natural Science Foundation of China (NSFC) (Grants No. 12374160) and
the Natural Science Foundation of Shandong Province (Grants No. 2023HWYQ-009).}

\end{document}